\documentclass[graybox]{svmult}

\usepackage{type1cm}        
\usepackage{makeidx}         
\usepackage{graphicx}        
\usepackage{multicol}        
\usepackage[bottom]{footmisc}

\makeindex             

\begin{document}

\title*{On configuration space, Born's rule and ontological states}
\author{Hans-Thomas Elze}
\institute{Hans-Thomas Elze \at Dipartimento di Fisica ``Enrico Fermi'', \\ \noindent Universit\'a di Pisa, Largo Pontecorvo 3, I-56127 Pisa, Italia 
\\ \noindent 
\email{elze@df.unipi.it}}


%
%
\maketitle

\abstract{It is shown how configuration space, possibly encompassing ordinary spatial structures, Born's rule, and ontological states aiming to address an underlying reality beyond Quantum Mechanics relate to each other in models of Hamiltonian cellular automata.}

\section*{}  
\begin{svgraybox}
This paper is dedicated to the memory of Walter Greiner.  -- Always Walter shared 
freely his wise suggestions concerning topics worth exploring in physics and beyond, all the way to `career moves' to be done by his students. Yet, rarely did I follow his fatherly advice. -- 
It happened now and then during the weekly ``Palaver'' seminars that Walter quickly dismissed foundational or interpretational issues as irrelevant for physics, 
no matter who dared to mention them. 
-- Yet, quite recently, following my talk at Walter's {\it F}ranfurt {\it I}nstitute of {\it A}dvanced {\it S}tudy, about new approaches to understand Quantum Mechanics 
as emerging from something `intelligible' beneath, he encouraged me strongly to pursue these ideas \dots  
\end{svgraybox}

\section{Introduction}
\label{sec:1}
It shall {\it not} be our concern to derive Quantum Mechanics (QM) from somehow physically motivated and more or less parsimonious sets of axioms, such as recent information theoretical {\it reconstructions} of QM \cite{DAriano,Hoehn1,Hoehn2,Hoehn3}, nor to propose yet  another {\it interpretation} of QM, of which there are already too many (Copenhagen, Many Worlds, Qbism, \dots )  to sort out seemingly incompatible aspects \cite{interpretations}. We will explore {\it deformations of quantum mechanical models}, due to the presence of a finite discreteness scale $l$ (of length or time, choosing $c=\hbar =1$, henceforth). And what can be learnt from them regarding basic concepts of QM. 

Presently, we shall reconsider 
{\it configuration space} and the {\it Born rule} in relation to the foundational hypotheses   introduced recently by Gerard 't\,Hooft, 
in particular, the existence and relevance of {\it ontological states} of cellular automata that may give rise to the  reality described by physics \cite{tHooftBook}. 
 
Our discussion will refer to the class of {\it Hamiltonian Cellular Automata} (CA) \cite{Elze1}. See Ref.\,\cite{Elze3} for a wider perspective on striving for the understanding of QM as an emergent structure, the {\it raison d'\^etre} for Hamiltonian CA, and  numerous references to earlier work. 

\section{Hamiltonian Cellular Automata -- some essentials} 
\label{sec:2}
The {\it Hamiltonian CA} describe discrete linear dynamical  systems that show quantum features --  especially, but not only in the continuum limit \cite{Elze1,Elze2}.  

These are deterministic classical CA with  denumerable  degrees of freedom (``bit processors''). The  state of such a CA is described by {\it integer valued} coordinates $x_n^\alpha$ 
and momenta $p_n^\alpha$, where  
$\alpha\in {\mathbf N_0}$ labels different degrees of freedom and  
$n\in {\mathbf Z}$ successive states. -- We mention that a generalization studied earlier  consists in introducing also a dynamical discrete time coordinate together with its associated  momentum, which must be distinguished from the `CA clock' time $n$ \cite{Elze1}. 

Only {\it finite differences} of variables can play a role in the dynamics here, where {\it no infinitesimals} or ordinary derivatives are available!

Dynamics and symmetries of such systems are all contained in a suitable {\it action principle} \cite{Elze1}, the consistency of which places severe constraints on its detailed form. 
Ultimately, this is responsible for the essential {\it linearity} of the QM models, which result in the continuum limit within this class of CA, as we have shown earlier. 

This action principle yields {\it finite difference} equations of motion: 
\begin{eqnarray}\label{eomx} 
\dot x_n^\alpha &=& S_{\alpha\beta}p_n^\beta +A_{\alpha\beta}x_n^\beta  
\;\;, \\ [1ex] \label{eomp}
\dot p_n^\alpha &=& -S_{\alpha\beta}x_n^\beta +A_{\alpha\beta}p_n^\beta  
\;\;, \end{eqnarray}  
with given integer-valued symmetric and antisymmetric matrices,
$\hat S\equiv \{S_{\alpha\beta}\}$  and $\hat A\equiv\{A_{\alpha\beta}\}$, respectively, 
defining the model under consideration (summation over repeated Greek indices applies, 
except where stated otherwise). Note that we introduced the notation 
$\dot O_n:=O_{n+1}-O_{n-1}$. -- This may guide the eye and indicate an analogy with Hamilton's equations in the continuum, suggesting the name {\it Hamiltonian  CA}; however, in distinction to usually first-order derivatives, we encounter second-order finite difference operators here.   

This leads to discrete time reversal invariance of the Eqs.\,(\ref{eomx})-(\ref{eomp}); 
{\it i.e.} the equations are invariant under reversal of the updating direction. The state of such a CA can be updated in both directions, 
$(n\mp 1,n)\;\rightarrow\;(n\pm 1)$, given appropriate initial values of its variables.   

The pair of equations of motion can be combined into one equation: 
\begin{equation}\label{discrSchroed}
\dot\psi_n^\alpha\;=\;-iH_{\alpha\beta}\psi_n^\beta
\;\;, \end{equation} 
together with its adjoint, by introducing the self-adjoint ``Hamiltonian'' matrix, 
 $\hat H:=\hat S+i\hat A$, and complex integer-valued ({\it Gaussian 
integer}) state variables, 
$\psi_n^\alpha :=x_n^\alpha +ip_n^\alpha$. 
This obviously looks like the 
Schr\"odinger equation, despite involving only Gaussian integer quantities.   

The resemblance between Eq.\,(\ref{discrSchroed}) and the Schr\"odinger equation in the continuum, for a whole class of models, can hardly be accidental. Indeed, we have constructed an invertible map between the discrete equation describing Hamiltonian CA in terms of the variables $\psi_n^\alpha$ and a continuous time equation describing the {\it same} CA in terms of a complex ``wave function''  $\psi^\alpha (t)$ \cite{Elze1}.  This has been achieved with the help of Sampling Theory \cite{Jerri,Shannon}, introducing the 
finite discreteness scale $l$ in terms of a bandwidth limit or high-frequency cut-off  for all such wave functions.  

Thus, we obtain the Schr\"odinger equation, yet modified to incorporate a power series in higher-order time derivatives with $l$-dependent coefficients: 
\begin{equation}\label{SchroedCont}
2\sinh (l\partial_t)\psi^\alpha (t)=-i\hat H_{\alpha\beta}\psi^\beta (t) 
\;\;. \end{equation} 
This implies corresponding modifications of other standard results of QM and may have interesting phaenomenological consequences.  

Naturally, known features of QM are reproduced in the continuum limit, $l\rightarrow 0$, of the {\it deformation} of QM implied here \cite{Elze2,Elze3}.

\section{Conservation laws, configuration space and a glimpse of the Born rule} 
\label{sec:3}   
There are  $l$-dependent conservation laws in one-to-one correspondence with those of the corresponding  quantum mechanical model obtained for $l\rightarrow 0$. 
With the help of Eq.\,(\ref{discrSchroed}) and its adjoint one easily verifies the theorem:  
\begin{itemize} 
\item For any matrix $\hat G$ that commutes with 
$\hat H$, $[\hat G,\hat H]=0$, there 
is a {\it discrete conservation law}: 
\begin{equation}\label{Gconserv} 
 \psi_n^{\ast\alpha}G_{\alpha\beta}\dot\psi_n^\beta +
\dot\psi_n^{\ast\alpha}G_{\alpha\beta}\psi_n^\beta =0 
\;\;. \end{equation}  
For self-adjoint $\hat G$,  defined in terms of Gaussian integers, 
this statement concerns real integer quantities.  
\item   Rearrangement of Eq.\,(\ref{Gconserv}) gives the related  
{\it conserved quantity} $q_{\hat G}$: 
\begin{equation}\label{qG} 
q_{\hat G}:=\psi_n^{*\alpha}\hat G_{\alpha\beta}\psi_{n-1}^\beta+\psi_{n-1}^{*\alpha}\hat G_{\alpha\beta}\psi_n^\beta 
=\psi_{n+1}^{*\alpha}\hat G_{\alpha\beta}\psi_n^\beta +\psi_n^{*\alpha}\hat G_{\alpha\beta}\psi_{n+1}^\beta   
\;\;, \end{equation}  
{\it i.e.} a complex (real  for $\hat G=\hat G^\dagger$) integer-valued correlation function which is invariant under a shift $n\rightarrow n+m$, $m\in\mathbf{Z}$. 
\item For $\hat G:=\hat 1$, the conservation law states a constraint on 
the state variables: 
\begin{equation}\label{normal}  
q_{\hat 1}=2\mbox{Re}\;\psi_n^{*\alpha}\psi_{n-1}^\alpha
=2\mbox{Re}\;\psi_{n+1}^{*\alpha}\psi_n^\alpha =\mbox{const}   
\;\;. \end{equation}  
\end{itemize} 

The latter replaces for discrete CA the familiar {\it normalization}  
of state vectors in QM, to which it reduces in the limit $l\rightarrow 0$. This follows from mapping the discrete to the continuum version by applying Shannon's reconstruction theorem, to which we alluded above \cite{Elze1}. Here, we obtain:   
\begin{eqnarray}\label{Qcont1}  
\mbox{const}=q_{\hat 1}
&=&\mbox{Re}\;\psi^*(t)\cosh\big [l\frac{\mbox{d}}{\mbox{d}t}\big ]\psi (t)
\\ [1ex] \label{Qcont2} 
&=&\psi^{*\alpha}(t)\psi^\alpha (t)
+\frac{l^2}{2}\mbox{Re}\;\psi^{*\alpha}(t)\frac{\mbox{d}^2}{\mbox{d}t^2}\psi^\alpha (t) 
+\mbox{O}(l^4)   
\;\;, \end{eqnarray} 
displaying the $l$-dependent corrections to the continuum limit, namely the normalization $\psi^{*\alpha}\psi^\alpha =\mbox{const}$, which is usually conserved but not in the present case of CA. 

In order to interpret the {\it conserved `two-time' correlation} function $q_{\hat 1}$ of Eq.\,(\ref{normal}),  we recall that the greek indices abstractly label different CA {\it degrees of freedom}.  These could relate to an internal space or to localized sites of a spatial structure. However, the entire formalism describing CA and demonstrating their quantum features, so far, is independent of attaching such a traditional meaning to the degrees of freedom. This observation is valid for  multipartite CA as well \cite{Elze2,Elze3}. In this sense, the {\it primacy of configuration space} appears naturally 
here.\footnote{Mathematically speaking, the Gaussian integer wave functions $\psi$, with components $\psi^\alpha ,\;\alpha\in {\mathbf N_0}$, can be seen as elements of a linear space endowed with an
integer-valued scalar product, $\psi^*\psi :=\psi^{*\alpha}\psi^\alpha$,  {\it i.e.} a unitary space. Taking its incompleteness into account, the space of states can be classified as a
pre-Hilbert module over the commutative ring of Gaussian integers \cite{Elze2}.} 
\vskip 0.2cm 
This will not lead us to immediate practical consequences or new theoretical insights, 
however, we speculate that the physics 
of matter {\it in} space or {\it in} spacetime and  the physics {\it of} space or  {\it of} spacetime itself emerge together from underlying discrete structures and CA like processes (possibly of a kind totally unexpected). 
Loosely speaking, some evolving complex aggregates of CA degrees of freedom will show up as matter and some as space(time) and must be linked in intricate ways.   

Experiencing effects of these phenomena only at scales some nineteen or more orders of magnitude away from the Planck scale, say, QM plays the role of an effective description that has been found most successful when applied to atomic or subatomic particles and the forces that influence their motion. This should include, in principle, the  description of macroscopic amounts of solid, liquid, or gaseous matter. 

However, since the early days of QM doubts shroud the beauty of this  picture  where it is supposed to reflect, within QM, the whole process of a measurement  undertaken in a laboratory or elsewhere (let alone the alive physicist involved). There has been  
a puzzling gap in understanding this  {\it measurement problem}, unless one is willing to undersign one of the available  ``interpretations'' of QM which eliminate this puzzle, at the expense of introducing others \cite{interpretations}  (for an overview 
and proposal of a solution, see \cite{Nieuwenh}). 
We shall come back to this issue when discussing ontological CA states. 
\vskip 0.2cm 
Besides that Eq.\,(\ref{normal}) replaces the normalization of state vectors, a necessary ingredient of the {\it Born rule} in QM, we can rephrase its content by relating it to a  {\it counting procedure} as follows. 

Recall that 
$\psi_n^\alpha :=x_n^\alpha +ip_n^\alpha$. Thus, 
for a pair of successive states labelled by $n$ and $n+1$ and for each degree of freedom $\alpha$, there is a pair of integer components $x_n^\alpha$ and $p_n^\alpha$ that may be  called the (numbers of) $x^\alpha$- and $p^\alpha$-{\it initials}, respectively. Correspondingly, the 
integer components $x_{n+1}^\alpha$ and $p_{n+1}^\alpha$ are called the (numbers of) $x^\alpha$- and $p^\alpha$-{\it finals}, respectively. This includes the possibility of zero or missing ({\i.e.} negative) numbers of initials or finals. Summing the product of 
$x^\alpha$-initials and -finals and the product of $p^\alpha$-initials and -finals 
defines the number of $\alpha$-{\it links}, $L_\alpha$: 
\begin{equation}\label{Lalpha}
L_\alpha := x_{n+1}^{*\alpha}x_n^\alpha +p_{n+1}^{*\alpha}p_n^\alpha 
\;\;\in\mathbf{Z}\;\;, \end{equation}
dropping the summation convention henceforth. Then, the total number of links is given by 
$L:=\sum_\alpha L_\alpha$ and Eq.\,(\ref{normal}) states: 
\begin{equation}\label{Lconserved}
L=q_{\hat 1}/2=\mbox{const} 
\;\;, \end{equation} 
{\it i.e.}, the {\it link number is conserved}. -- The effect of the CA evolution according to the equations of motion (\ref{eomx}) and (\ref{eomp}) is to change or  
redistribute the numbers of initials and finals over the available degrees of freedom, while keeping the total number of links constant.\footnote{One may ponder the possibility that 
the counting of links and their conservation law could be generalized such that $\alpha$-initials are linked to a {\it finite set} of $\alpha '$-finals, including the $\alpha$-finals. This could be pictured as a {\it forward lighcone}-like structure, considering  spatial sites 
labelled by index $\alpha$ etc. Could there correspondingly exist Hamiltonians that would conform with {\it locality} in the usual special relativistic sense?}   

The  relative {\it weights}, 
$w_\alpha :=L_\alpha /L\;(L\neq 0)$, with $\sum_\alpha w_\alpha =1$, generally, are not confined to $[0,1]$. However, they 
correspond uniquely 
to the QM probabilities $p_\alpha :=\psi^{*\alpha}(t)\psi^\alpha (t)/\psi^*\psi$, 
since $w_\alpha\rightarrow p_\alpha$ for $l\rightarrow 0$, in accordance with Eqs.\,(\ref{normal})--(\ref{Qcont2}).\footnote{The possibility of {\it negative} link numbers $L_\alpha$ or weights $w_\alpha$ reminds of ``probabilities'' falling outside of 
$[0,1]$ appearing in QM as discussed by Wigner, Feynman, and others; see, {\it e.g.}, the  reviews by Khrennikov \cite{Khrennikov} and by M\"uckenheim {\it et al.} \cite{Mueckenheim}} -- In distinction, the case $L=0$ does not allow a meaningful continuum 
limit. We will encounter an example and its interpretation in Sec.\ref{sec:4}, 
introducing {\it ontological states}. 

Thus, we recover attributes of the {\it Born rule}. 
Of course, our description of deterministic Hamiltonian CA, so far, says as little as quantum theory about the origin of the {\it randomness   
of  experimental outcomes}. However, the hypothesis 
of ontological states and their generally statistical relation to the (pre-)quantum states of CA will add a new element changing this situation. 

\section{The hypothesis of ontological states}
\label{sec:4}  
Ontological states that underlie quantum and, {\it a fortiori}, classical states of physical objects are central to the {\it Cellular Automaton Interpretation} (CAI) of QM \cite{tHooftBook}.  

There is ample motivation to reexamine the foundations of quantum theory in perspective of essentially classical concepts -- above all, determinism and existence of ontological states of reality -- which stems from observations of quantum features in a large variety of deterministic and, in some sense, ``classical'' models \cite{Elze1,Elze2,Elze3}. -- It is worth emphasizing that quantum states here are considered to form part of the mathematical language used, they are ``templates'' for the description of the ``reality beneath'', including ontological states and their deterministic dynamics. 

Finite and discrete CA may provide the necessary versatility to accommodate ontological states and their evolution, besides (proper) quantum ``template'' states (especially in the continuum limit). The following general remarks serve to obtain an operational understanding of what we are looking for in such models:   
\vskip 0.2cm\noindent 
{\it ONTOLOGICAL STATES} ($\cal OS$) are states a deterministic physical system can be in. They are denoted by    
$|A\rangle ,\; |B\rangle ,\; |C\rangle ,\;\dots\;$. The set of all states may be very large, but is assumed to be denumerable, for simplicity.
\vskip 0.1cm\noindent 
There exist {\it no superpositions} of $\cal OS$ ``out there'' as part of physical reality. 
\vskip 0.1cm\noindent  
The $\cal OS$ evolve by {\it permutations} among themselves,  
$\dots\rightarrow |A\rangle\rightarrow |D\rangle\rightarrow |B\rangle\rightarrow\;\dots\;,$ for example. Apparently this is the only possibility, besides  producing a growing set of states or superpositions, which do not belong to the initial set of $\cal OS$.  
\vskip 0.1cm\noindent 
By declaring the $\cal OS$ to form an orthonormal set, fixed once for all, a {\it Hilbert space} can be defined.  -- Operators which are diagonal on this set of 
$\cal OS$ are {\it beables}. Their eigenvalues describe physical properties of the $\cal OS$, corresponding to the 
abstract labels $A,\; B,\; C,\;\dots\;$ above.   
\vskip 0.2cm\noindent  
{\it QUANTUM STATES} ($\cal QS$) are superpositions of $\cal OS$. These are   
templates for doing  physics with the help of mathematics. --  
The amplitudes specifying superpositions need to be interpreted, when applying the formalism to describe experiments. 
Here the {\it Born rule is built in},  {\it i.e.}, by definition! By experience, interpreting amplitudes in terms of probabilities has been an amazingly useful invention.\footnote{It is indeed possible to change the proportionality  between {\it absolute values squared} of complex amplitudes and probabilities into a  more complicated relation, however, only at the expense of mathematical simplicity \cite{tHooftBook}.}  
\vskip 0.2cm
From here the machinery of QM can be seen to depart, incorporating especially the powerful techniques related to unitary transformations. The latter exist, in general, only in a rudimentary discrete form on the level of $\cal OS$, due to the absence of superpositions. 

While quantum theory has been very effective in describing experiments, its {\it linearity} is the characteristic feature of the unitary dynamics embodied in the Schr\"odinger 
equation.\footnote{This linearity is reflected by the Superposition Principle and  entails interference effects and the possibility of nonclassical correlations among parts of  composite objects, {\it i.e.} entanglement in multipartite systems.} For a prospective ontological theory, it is of interest that QM  remains notoriously  indifferent to any reduction or collapse process one might be tempted to add on, in order to modify the collapse-free linear evolution and, in this way, solve the measurement problem.

Concluding this brief recapitulation of some essential points of CAI, in particular how QM fits into this wider realistic picture, one more remark is in order, concerning the {\it absence of the measurement problem}: 
\vskip 0.2cm\noindent  
{\it CLASSICAL STATES} of a macroscopic deterministic system, including billiard balls, pointers of apparatus, planets, are {\it probabilistic distributions} of $\cal OS$, since  any kind of repeated experiments performed by physicists, with only limited control of the circumstances, pick up different initial conditions regarding the $\cal OS$. Hence, different outcomes of apparatus readings must generally be expected. Yet any reduction or collapse to a $\delta$-peaked distribution, say, of pointer positions is only an apparent effect, induced by the intermediary use of quantum mechanical  templates in describing the evolution of $\cal OS$ during an experiment. Ontologically speaking, there were/are no superpositions, to begin with, which could possibly collapse \cite{tHooftBook}!
\vskip 0.2cm 
This provides a strong motivation for pursuing an approach to understand reality as based on ontological states.  

\subsection{Avoiding superpositions}
\label{subsec:4.1}
We recall that  $\cal OS$ evolve by permutations among themselves. This is  quite different from the behaviour commonly found in QM, namely the dynamical formation of {\it superposition states} (except for stationary states). 

In order to assess the formation, or not, of superpositions by an evolving Hamiltonian CA, we first remind ourselves that the Schr\"odinger equation is formally solved by:  
$$\partial_t\psi(t)=-i\hat H\psi(t)\;\;\;\Rightarrow\;\;\;\psi(t)=\mbox{e}^{-i\hat Ht}\psi (0)\;\;,$$ 
given the initial state $\psi (0)$, and consider the analogous formal solution of the CA equation of motion (\ref{discrSchroed}),  
$\dot\psi_n=\psi_{n+1}-\psi_{n-1}=-i\hat H\psi_n$. 
In terms of an auxiliary operator $\hat\phi$, defined by $2\sin\hat\phi :=\hat H$, one finds indeed: 
\begin{equation}\label{discrsol}
\psi_n=\frac{1}{2\cos\hat\phi}\big (\mbox{e}^{-in\hat\phi}[\mbox{e}^{i\hat\phi}\psi_0+\psi_1]
+(-1)^n\mbox{e}^{in\hat\phi}[\mbox{e}^{-i\hat\phi}\psi_0-\psi_1]\big )
\;\;. \end{equation}  
where {\it two initial states} are required, $\psi_0$ and $\psi_1$, corresponding to  the fact that  
Hamiltonian CA are described by a {second-order finite difference equation}.  

With the help of the general solution (\ref{discrsol}), one obtains: 
\begin{equation}\label{Tsol1}
\psi_n=\hat T(n-m+1)\psi_{m+1}+\hat T(n-m)\psi_{m}  
\;\;, \end{equation} 
where $\hat T$ is a transfer operator that can be read off by comparing with the explicit form 
of this relation. This generalizes the composition law for the unitary time 
evolution operator in QM. -- Furthermore, the simple exponential expression for the solutions in QM can be  
recovered from Eq.\,(\ref{discrsol}) by taking the appropriate limits    
$n\rightarrow\infty$ and $l\rightarrow 0$, keeping $n\cdot l$ fixed,  {\it and} choosing initial conditions such that 
$\psi_1\equiv\psi_0$. In this case, we have:    
\begin{equation}\label{Tsol2} 
\psi_n=\big [\hat T(n+1)+\hat T(n)\big ]\psi_0
\;\;. \end{equation}     
The Eq.\,(\ref{Tsol1}) and especially Eq.\,(\ref{Tsol2}) tell us to generally expect the formation of superposition states and, therefore, {\it not} ontological states which evolve by permutations among themselves.  

While this seems to severely obstruct the search for $\cal OS$ from the outset,   
we now present a first simple example illustrating that evolving   
$\cal OS$ are possible in a  {\it two-state} CA. -- 
Consider the CA described by $\psi_n^\alpha ,\;\alpha =1,2\;$,  with equation of motion given by:  
\begin{equation}\label{exEoM}
\psi_n=\psi_{n-2}-i\hat H_2\psi_{n-1}\;,\;\;
\psi_n\equiv \left (\begin{array}{c}\psi_n^1 \\ \psi_n^2\end{array}\right )\;,\;\; 
\hat H_2:=\left (\begin{array}{c  c} 0 & 1 \\ 1 & 0 \end{array}\right )\equiv\hat\sigma_1
\;\;. \end{equation}   
Furthermore, we choose two orthogonal initial states, 
$\psi_0=(1,0)^t$ and $\psi_1=(0,1)^t$. By solving the equation of motion most simply 
by iteration, we obtain the following sequence of states:  
\begin{eqnarray}\label{twoditer}
&\;&\psi_0\;,\;\psi_1\;,\;\psi_2=(1-i)\psi_0\;,\;\psi_3=-i\psi_1\;,
\nonumber \\ [1ex] 
&\;&\psi_4=-i\psi_0\;,\;\psi_5=-(1+i)\psi_1\;,\;
\psi_6=-\psi_0\;,\;\psi_7=-\psi_1\;,\;\dots 
\;\;, \end{eqnarray} 
which after four more steps begins to reproduce the initial pair of states. 
 
Here the normalization of the states, considered as if of vectors embedded in a Hilbert space for a moment, changes dynamically. This would be a disaster in QM!    
However, for Hamiltonian CA this norm is not conserved. Instead, it is replaced by a {\it conserved `two-time' correlation function}, {\it cf.} Eqs.\,(\ref{qG})--(\ref{normal}),   
reproducing the norm conservation only in the continuum limit \cite{Elze1}. 

Thus, apart from the change of normalization, the evolution here essentially swaps 
two orthogonal input states, once per updating step. This provides a very simple example of a  CA evolving $\cal OS$, in agreement with CAI.    
 
\subsection{More interesting CA and their $\cal OS$} 
\label{subsec:4.2}
Following the primitive example just given, the question arises, whether there exists any  
generalization describing something more interesting.  

Besides systems with block diagonal $\hat H$ for multiple two-state components, 
one may try a higher-dimensional state space for generalizations of the model of Eqs.\,(\ref{exEoM})--(\ref{twoditer}). 
Indeed, we find easily that the Hamiltonians: 
\begin{equation}\label{H3H4}
\hat H_3:=\left (\begin{array}{c  c c} 0 & -i & 1 \\ i & 0 & -i \\ 1 & i & 0 \end{array}\right ) \;\;,\;\;\;
\hat H_4:=\left (\begin{array}{c  c c c} 0 & -i & 0 & 1 \\ i & 0 & -i & 0 \\ 0 & i & 0 & -i \\ 1 & 0 & i & 0 \end{array}\right )
\;\;, \end{equation} 
for three- and four-state CA, respectively, lead to analogous evolution-by-permutation of $\cal OS$ as the previous example, $\hat H_2$ of Eq.\,(\ref{exEoM}). We may generally consider the $m$-dimensional state space with Hamiltonian:  
\begin{equation}\label{Hm}
\hat H_m:=\left (\begin{array}{c c c c c c} 
 0 & -i & 0 & \cdots & 0 & 1 \\ 
  i & 0 & -i & 0 & \cdots & 0 \\ 
    &    &     &    &               &    \\
    &    &  & \ddots &  &  \\ 
 0 & \cdots & 0 & i & 0 & -i \\ 
 1 & 0 & \cdots & 0 & i & 0 
\end{array}\right )
\;\;, \end{equation} 
which works like the previous examples for `neighbouring' pairs of orthogonal 
initial states,    
$\psi_0=\psi^{(k)}:=(0,\dots ,0,1,0,\dots ,0)^t$, with nonzero $k$-th entry ($1\leq k\leq m-1$),  
and correspondingly $\psi_1=\psi^{(k+1)}$. 

The CA evolution with Hamiltonian $\hat H_m$, Eq.\,(\ref{Hm}), does not change the normalization of these states, but can introduce phases ($\pm1,\pm i$) when permuting them. To give an explicit example, choosing  
$\psi_0=\psi^{(m-1)}$ and $\psi_1=\psi^{(m)}$, the result of one updating step is 
$\psi_2=\psi_0-i\hat H_m\psi_1=-i\psi^{(1)}$. Such phases are carried on by further updating steps until they are eventually cancelled and the initial configuration reappears, only after $4m$ updates.   

Therefore, all the $4m$ $\cal OS$, which eventually {\it differ by phases}, must be considered as {\it different states} here. 
Note that in 
$x'+ip'\equiv\psi'=i\psi \equiv i(x+ip)$, for example, the roles of coordinates (real parts) and momenta (imaginary parts) are exchanged, $x'=-p$, $p'=x$. Such states cannot be seen as embedded in a projective space, which would be the case of 
normalized states in QM. Loosely speaking, they are `more classical'.

All pairs of (initial) states of this kind have the conserved {\it link number} $L=0$, cf. Eqs.\,(\ref{Lalpha})--(\ref{Lconserved}). Whereas initial configurations with 
$\psi_0\equiv\psi_1$ have $L>0$ -- they could serve as quantum mechanical templates (in the continuum limit), {\it cf.} introduction of this Sec.\ref{sec:4} and discussion of Eqs.\,(\ref{Tsol1})--({\ref{Tsol2}). Thus, the conserved link number illustrates the {\it ontology conservation law} \cite{tHooftBook}. 

The dynamics described by $\hat H_m$ resembles the {\it cogwheel model} discussed in 
Refs.\,\cite{Elze4,ElzeSchipper,tHooftBook} -- first introduced by 't\,Hooft as a `particle'  making uniform jumps over fixed positions on a circle, one per fixed time interval -- which has been shown to have surprising quantum features (providing a discrete representation of the quantum harmonic oscillator).  However, while evolution was given by a unitary first-order updating rule in those models, it is a second-order process determined by a self-adjoint operator, $\hat H_m$, in the present case.  
\vskip 0.2cm 
Before closing, a remark is in order here. Namely, {\it interacting multipartite Hamiltonian CA} \cite{Elze3}, in particular those consisting of two-state ``Ising spin'' subsystems, offer an alternative to look for more complex behaviour concerning ontological states than in the one-component examples chosen here for simplicity. Such systems 
have been considered recently \cite{Elze4}, with further results to be presented elsewhere. Generalization in this direction seems necessary, in order to develop ontological models that possibly can serve as a realistic base from which the theory of interacting relativistic quantum fields can emerge in analogous ways as the QM models we described. 

\section{Conclusion} 
\label{sec:5}
In retrospect, one could subsume our results, {\it cf.} especially the summary given in Sec.\ref{sec:2}, as pertaining to a particular {\it discretization} of QM, which introduces the finite scale $l$, conceivably the Planck scale. 

However, we have reported in Sec.\,\ref{sec:3} the resulting conservation laws in such models of discrete {\it Hamiltonian CA}, which are entirely described in terms of integer valued quantities, and illustrated the one-to-one correspondence with those of continuum models of quantum theory, which are recovered for $l\rightarrow 0$. In particular, the conserved quantity, a two-time correlation, replacing here the conserved norm of a QM state vector, has led us to a simple interpretation in terms of a {\it counting procedure} -- unnoticeable in quantum theory, since there only the coincidence limit of the correlation matters!\footnote{Similar observations hold for all conservation laws and could be of phenomenological interest.}  This seems to shed a different light on the 
{\it Born rule}. 

Up to this point, though, we still did not encounter the randomness eventually seen in experimental outcomes. 

As we have argued, following 't\,Hooft, the {\it probabilistic features of QM} can be understood to result from the available mathematical description of the underlying deterministic reality. The unavoidable mismatch between the two can be precisely traced to the {\it nonexistence of superposition states} of ``stuff'' that is ontologically there and the powerful use that is made of such formal superpositions in quantum theory. 

By the {\it hypothesis of 
ontological states}  and by illustrating their existence within the present class of discrete models, of Hamiltonian CA kind,  
one has left standard QM, as suggested by the {\it Cellular Automaton Interpretation} \cite{tHooftBook}, {\it cf.} the introduction to Sec.\ref{sec:4}.  This may provide some indication that reality can be understood to exist ``out there'', sometimes misnamed ``Einstein's dream'', that ontological states describe states in which a deterministic physical system can be and how it evolves. Yet QM is neither abandoned nor has quantum theory been changed, but one begins to understand it as a most effective mathematical 
construct/language to describe the reality of what we perceive. Much more is left to be done. 

\begin{acknowledgement}
It is a pleasure to thank Dirk Rischke and Horst St\"ocker for the invitation to the 
{\it International Symposion on Discoveries at the Frontiers of Science} in honour of Walter Greiner 
(FIAS, Frankfurt, June 26-30, 2017), for support, and especially for kind hospitality. 
\end{acknowledgement}
%

%
%
%

\end{document}